\documentclass[12pt]{article}

\usepackage[margin=1in]{geometry}
\usepackage[utf8]{inputenc}
\usepackage{color,url}
\usepackage[authoryear]{natbib}
\usepackage{amssymb}
\usepackage{wrapfig}
\usepackage{graphicx}
\usepackage{pifont}
\usepackage{hyperref}

\newcommand{\xmark}{\ding{55}}%

\newcommand{\wontfix}{\rlap{$\square$}{\large\hspace{1pt}\xmark}}

\begin{document}

\noindent {\LARGE{{\textbf{Astro2020 Science White Paper}}}}\\
\vspace{1cm}

\noindent {\LARGE{{\textbf{Mapping Cosmic Dawn and Reionization:\vspace{0.1cm}\\Challenges and Synergies}}}} \\
\vspace{0.1cm}

\noindent \textbf{Thematic Areas:} \hspace*{60pt} $\square$ Planetary Systems \hspace*{10pt} $\square$ Star and Planet Formation \hspace*{20pt}\linebreak
$\square$ Formation and Evolution of Compact Objects \hspace*{31pt} \wontfix Cosmology and Fundamental Physics \linebreak
  $\square$  Stars and Stellar Evolution \hspace*{1pt} $\square$ Resolved Stellar Populations and their Environments \hspace*{40pt} \linebreak
  \wontfix    Galaxy Evolution   \hspace*{45pt} $\square$             Multi-Messenger Astronomy and Astrophysics \hspace*{65pt} \linebreak
\vspace{0.1cm}

\noindent {\large{\textbf{Principal Authors:}}}\\ 
Marcelo A. Alvarez\footnote{\noindent University of California, Berkeley; marcelo.alvarez@berkeley.edu}, 
Anastasia Fialkov\footnote{\noindent University of Sussex; a.fialkov@sussex.ac.uk}, Paul La Plante\footnote{\noindent University of Pennsylvania; plaplant@sas.upenn.edu} \\
\vspace{0.1cm}

\noindent {\large{\textbf{Co-authors:}}}\\ 
James Aguirre, Yacine Ali-Ha\"imoud, George Becker, Judd Bowman, Patrick Breysse, Volker Bromm, Philip Bull, Jack Burns, Nico Cappelluti, Isabella Carucci, Tzu-Ching Chang,  Kieran Cleary, Asantha Cooray, Xuelei Chen, Hsin Chiang, Joanne Cohn, David DeBoer, Joshua Dillon, Olivier Dor\'e, Cora Dvorkin, Simone Ferraro, Steven Furlanetto, Bryna Hazelton, J. Colin Hill, Daniel Jacobs, Kirit Karkare, Garrett K. Keating, L\'eon Koopmans, Ely Kovetz, Adam Lidz, Adrian Liu, Yin-Zhe Ma, Yi Mao, Kiyoshi Masui, Matthew McQuinn, Jordan Mirocha, Julian Mu\~noz, Steven Murray, Aaron Parsons, Jonathan Pober, Benjamin Saliwanchik, Jonathan Sievers, Nithyanandan Thyagarajan, Hy Trac, Alexey Vikhlinin, Eli Visbal, Matias Zaldarriaga\\


\vfill 
\noindent {\large{\textbf{Abstract:}}}\\
Cosmic dawn and the Epoch of Reionization (EoR) are among the least explored observational eras
in cosmology: a time at which the first galaxies and supermassive black holes
formed and reionized the cold, neutral Universe of the post-recombination
era. With current instruments, only a handful of the brightest galaxies and
quasars from that time are detectable as individual objects, due to their
extreme distances. Fortunately, a multitude of multi-wavelength intensity
mapping measurements, ranging from the redshifted 21 cm background in the radio
to the unresolved X-ray background, contain a plethora of synergistic
information about this elusive era. The coming decade will likely see direct
detections of inhomogenous reionization with CMB and 21 cm observations, and a
slew of other probes covering overlapping areas and complementary physical
processes will provide crucial additional information and cross-validation. To
maximize scientific discovery and return on investment, coordinated survey
planning and joint data analysis should be a high priority, closely coupled to
computational models and theoretical predictions.

\vspace{2cm}
\thispagestyle{empty}

\newpage \setcounter{page}{1}
\subsection*{The High-redshift Frontier }
\vspace{-7pt}

The epochs of cosmic dawn and reionization are filled with opportunity for
exploring cosmological evolution.
Thus far, sensitive ground and space based instruments -- covering a wide range
of the electromagnetic spectrum -- have been used to detect, study, and catalog
individual objects well into the EoR, out to redshifts as high
as $z \sim 10$
\citep{2015ApJ...811..140B,2017ApJ...851L...8V,2018MNRAS.479.1055B}. However,
the bright objects studied with these instruments are not a representative
sample.
Without alternative probes of this elusive chapter of
cosmic history, significant gaps will remain.

Fortunately a complementary approach exists, focused on nearly uniform surveys of large regions of the sky over a broad range of wavelengths to obtain maps of the combined emission, scattering and
absorption from the stars, black holes, dust, and gas that existed in the early
universe. Such maps represent an unbiased sample of all the photons available
over the wavelengths and areas surveyed, and contain valuable information even
if all the discrete sources are confusion-limited, the value at each pixel
is noise-dominated, or both. A wide variety of map-based analysis techniques, many of
which have already been developed to study diffuse Galactic structure and the
cosmic microwave background (CMB), are available to extract the relevant
information.

In this white paper, we describe the promise and challenges of multi-wavelength
mapping of the high redshift universe in the coming decade. We argue that the
planned and proposed high-redshift surveys are highly synergistic -- the amount of
mutual information contained in data obtained using instruments with very
different systematic effects allows cross-validation of detections and highlights
information that is not apparently manifested in each individual
tracer. This synergistic approach might also be beneficial when extracting weak
signals hidden under strong foregrounds and systematics. Studying the aggregate
effects of the complex processes comes with
unique and significant challenges, both for astrophysical modeling in an era
about which we know very little, as well as for the mitigation of foreground
contamination and instrumental systematics. Consequently, significant
coordination between the various experiments that will gather the data used to
produce these maps is crucial, from instrument design, to survey strategy, and
finally joint analysis and cross-correlation.
\vspace{-5pt}

\subsection*{Intensity Mapping from the Radio to X-ray}
\vspace{-7pt}

The redshifted 21 cm background from the hyperfine transition of hydrogen is a probe of the neutral intergalactic medium
(IGM) and the history of its heating and reionization, while secondary CMB
anisotropies and spectral distortions from Thomson scattering by electrons
during reionization are sensitive to its overall morphology and history. Line
intensity mapping of various molecular and atomic lines can be used to
characterize properties of first stars and galaxies, and the unresolved X-ray
background contains information on the abundance and properties of energetic
high-redshift sources such as accreting black holes, supernovae, and X-ray binaries.\\

\noindent{\bf Redshifted 21 cm Background}\\
The 21 cm signal is produced by neutral hydrogen atoms in the medium outside the
star forming regions. The intensity and spatial fluctuations of this signal
depend on the ultraviolet and X-ray radiative backgrounds produced by the first
stars, galaxies, and black holes. Therefore, the signal can be used as a probe
of the early star and black hole formation.  21 cm cosmology is on the brink of
a revolution: the EDGES Low-Band collaboration has claimed the first detection
of an absorption feature from redshift $z \sim 17$ (yet to be confirmed,
\citealt{2018Natur.555...67B}), and pioneering global-signal experiments have
ruled out physically motivated models based on results from
$6 \lesssim z \lesssim 15$
\citep{2017ApJ...847...64M,2018ApJ...863...11M,2017ApJ...845L..12S,2018ApJ...858...54S}. Despite
the large variety of plausible theoretical scenarios (e.g.,
\citealt{2017MNRAS.472.1915C}), the first claimed detection does not comply with
standard astrophysical predictions.

Beyond this pioneering effort, the
international community is heavily committed to observations of the correlated
spatial fluctuations in the power spectrum of the 21 cm signal with facilities
such as: the Low Frequency Array (LOFAR), which yielded upper limits on the 21 cm
power spectrum at $z\sim 10$ \citep{2017ApJ...838...65P} and $20 \lesssim z \lesssim 25$
\citep{2018MNRAS.478.1484G}; the Hydrogen Epoch of Rreionization Array (HERA,
\citealt{deboer_etal2017}), which is projected to be fully constructed in late
2019; space-based missions such as DAPPer \citep{2018AAS...23231202B}; and the future Square Kilometre Array (SKA), whose Phase-1 observation is
expected to yield initial scientific results around 2023.
These power spectrum measurements can provide important results for testing the
reality of results from global signal measurements like EDGES
\citep{2019arXiv190202438F}.\\



\noindent{\bf Cosmic Microwave Background} \\
The ``patchy'' spatial distribution of electrons that result from what is very
likely to be a highly inhomogeneous reionization process is imprinted with high
fidelity through Thomson scattering of primary CMB photons.
The kinetic
Sunyaev-Zel'dovich (kSZ) effect is a secondary temperature anisotropy in the CMB
that is caused by the motion of electrons relative to the CMB rest frame and
interacting with CMB photons via Thomson scattering. The high-$z$ kSZ signal arises
primarily by the same patchy ionization field that generates the redshifted 21 cm
background fluctuations, so the CMB temperature and redshifted 21 cm background
should therefore be correlated. The magnitude and scale dependence of the kSZ
power spectrum are sensitive to the overall reionization history and
morphology. For example, the duration of the EoR affects the magnitude of the
kSZ signal, with longer epochs leading to larger kSZ signals.

Additionally, CMB secondary temperature and
polarization anisotropies are created by a modulation of primary temperature and
polarization fluctuations due to line of sight fluctuations in the optical depth
(``screening''; \citealt{2009PhRvD..79d3003D,2009PhRvD..79j7302D}) or the
generation of additional linear polarization in response to the local
quadrupolar anisotropy seen by electrons during reionization (``scattering'';
\citealt{2008ApJ...672..737M}). In
\citet{2009PhRvD..79d3003D}, a statistical technique for extracting the
inhomogeneous reionization signal from high-sensitivity measurements of the CMB
temperature and polarization fields was proposed. Recent work shows this signal
may be detectable by CMB-S4 through the B-mode power at
$50 \lesssim \ell \lesssim 500$ or by explicit reconstruction of the optical
depth at
the map level~\citep{2018JCAP...05..014R}.\\


\noindent{\bf Line Intensity Mapping (LIM)}\\
Characterizing properties of typical star forming high-redshift galaxies
requires measurements of molecular gas in objects which are below the detection
threshold of current telescopes. Molecular and atomic emission lines from dusty
high-redshift galaxies could be strong enough to allow intensity mapping (IM) of
this signal (for a recent review see \citealt{2017arXiv170909066K}).
The amplitude of the CO power spectrum was
recently tentatively detected in the shot noise regime by an intensity mapping
experiment, the CO Power Spectrum Survey \citep[COPSS, ][]{Keating2016}. This
experiment measured the power spectrum of CO for the $J_{1-0}$ transition with a
rest-frame frequency of 115.27 GHz from galaxies in the redshift range
$2.3 < z < 3.3$, an important proof-of-concept for line intensity
mapping of galaxies using CO. The CO Mapping Array Pathfinder (COMAP,
\citealt{2016AAS...22742606C}) experiment is currently underway and aims to
constrain the CO clustering power spectrum from galaxies at $2.4 < z < 3.4$ in
this first phase and at $6 < z < 8$ in a future phase. 

Metal fine structure lines probe gas in various environments and provide information about star formation rates in galaxies.
For
example [OIII] emission is relatively easy to model because it originates primarily from
ionized regions and was recently detected by ALMA at $7 \lesssim z \lesssim 9$, with
the most distant object located at $z = 9.11$
\citep{2018arXiv180600486H}. \citet{2018MNRAS.481L..84M} used high-resolution
simulations to study [OIII] emission from galaxies at $z > 7$ and suggest that
systematic surveys with JWST NIRCam can also probe the large-scale structure at
$z > 8$. The TIME-Pilot \citep{2014SPIE.9153E..1WC}, CONCERTO
\citep{lagache_2017}, and CCAT-prime \citep{2018SPIE10700E..5XP} experiments are specifically designed to observe [CII] emission lines from the
EoR and the post-reionization epoch.
CDIM \citep{2016arXiv160205178C} is a proposed satellite experiment that will
map the Ly$\alpha$ sky from redshifts $6 < z < 8$, and will contain valuable
information about the IGM from the EoR. The Ly$\alpha$ forest contains clues to
the detailed ionization and thermal state of the gas in the IGM, especially from
times post-reioinzation.\\

\noindent{\bf Unresolved X-ray Background}\\
Detection of X-ray emission from high redshifts could constrain
formation of super-massive black holes and properties of the first
population of X-ray binaries. Direct observations of high-$z$
populations require large integration times and can detect only the
brightest objects at high redshifts, while fainter and more common
high-$z$ sources are undetectable individually. As a result, the
observed objects are not a good statistical sample of the population,
and the clustering information is lost. A useful measurement is that
of the unresolved cosmic X-ray background (CXB) in the soft band
(0.5-2 keV) reported by the {\it Chandra} X-ray observatory, which
constrains both the luminosity and clustering of the most common
objects. Evidences and constrains on the X-rays emerging from the
EoR and observed in the CXB have been presented by
\citet{2012ApJ...752...46L},
\citet{2012MNRAS.427..651C,2013ApJ...769...68C}, and
\citet{2016ApJ...832..104M}. However, the origin of this cross-power is
uncertain. Future facilities will improve the situation.
A US-led \emph{Lynx} mission concept will have a somewhat smaller field
of view but much higher angular resolution and
sensitivity. \emph{Lynx} will be able to reach into extremely low flux
levels ($\sim 1/100$ of \emph{Chandra}) without being affected by
source confusion. As a result, the residual background fluctuations
due to undetected intrinsically faint $1<z<3$ objects are an order of
magnitude lower for \emph{Lynx} than for \emph{Athena} \citep{2018arXiv180909642T}.
\vspace{-7pt}

\subsection*{Synergies between Probes}
\vspace{-7pt}
High-redshift astrophysical processes should imprint consistent signatures that
naturally lead to correlations between the 21 cm signal, unresolved X-ray
background \citep{2017MNRAS.464.3498F,2018MNRAS.480...26M}, intensity mapping of
atomic
and molecular lines, and number counts of high-redshift galaxies. Although each of the probes discussed above provide important constraints on their own, their use in cross-correlation with the results from other probes can yield additional scientific insight. To enable synergies, coordinated surveys of the same part of the sky with different instruments is necessary. Measurements made from using cross-correlations potentially have a larger \textit{noise} than the auto-correlations, but are not \textit{biased}. They are also expected to be less sensitive to systematic errors.\\ 

\noindent{\bf CMB-21cm}\\ 
The 21 cm and kSZ signals should be non-vanishing
when computing the cross-correlation between them in map space. Because the kSZ
signal can be positive or negative (depending on whether the peculiar velocity
of the distant galaxies are toward or away from Earth), the simple
cross-correlation signal suffers from severe cancellation. One way forward is to
perform cross-correlations between the absolute value of the kSZ field, or the
value of the kSZ field squared \citep{ma_etal2018}. Alternatively, by computing
a higher-point statistic such as the bispectrum between two maps of the kSZ
field and one map of the T21 field, this cancellation can be
avoided. Accordingly, more detailed information about the EoR can be extracted
from the cross-correlation, with less contamination from signals at lower
redshift. Cross-correlation studies like these are also a way to confirm
properties of the EoR inferred from each individual experiment.\\

\begin{wrapfigure}{R}{0.5\textwidth}
  \centering
  \includegraphics[width=0.47\textwidth]{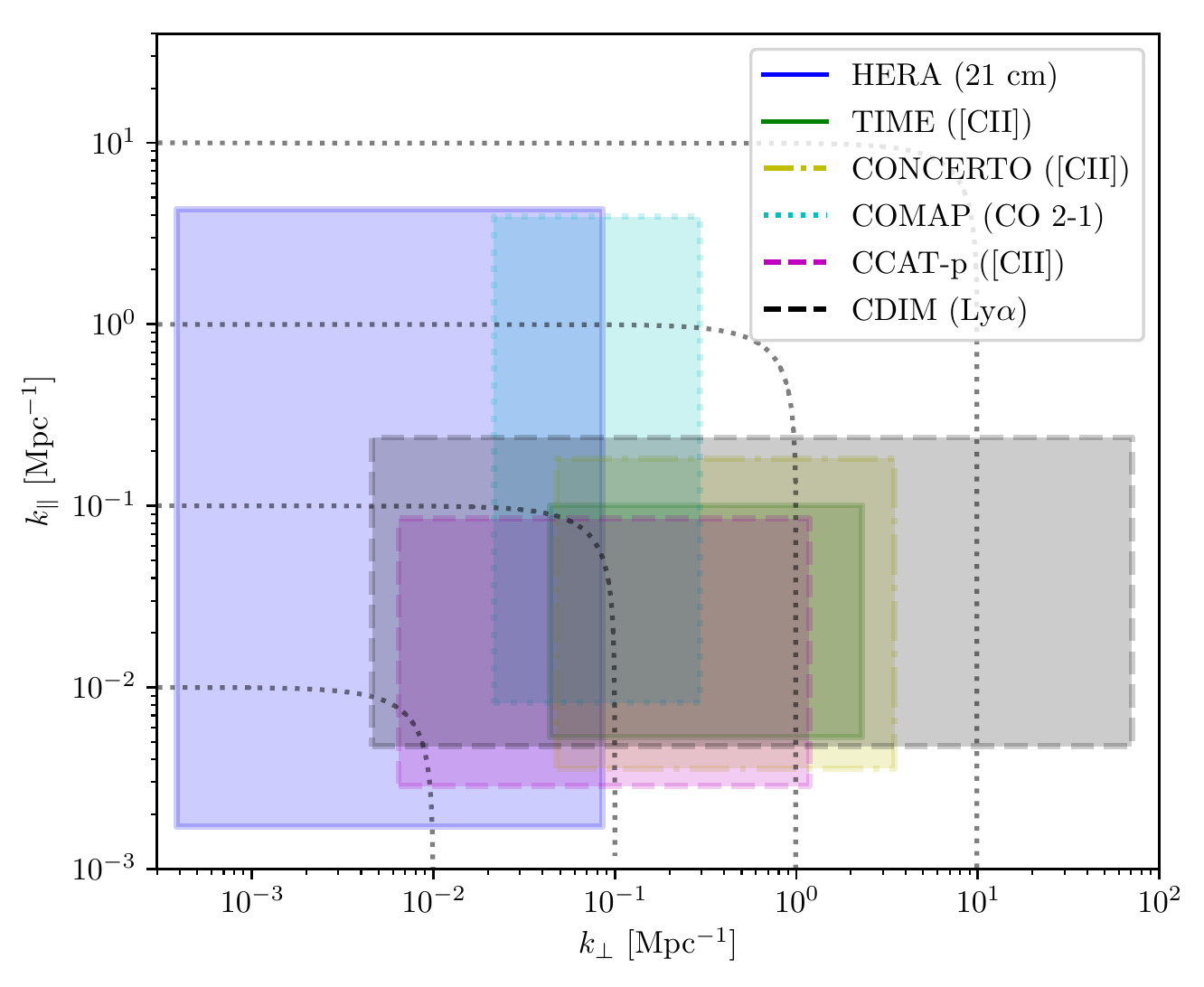}
  \caption{\small The overlap of various experiments in Fourier space as a function of comoving wavenumber in the plane of the sky ($k_\perp$) and along the line of sight ($k_\parallel$). Different boxes represent the rough coverage of various experiments. 
  Not captured in the figure is the redshift overlap of the various experiments -- in general, most of these cover the EoR and post-reionization epochs ($6 < z < 8$), which will provide for rich opportunities for cross-correlations.}
  \label{fig:kspace}
\end{wrapfigure}

\noindent{\bf LIM-21cm}\\
LIM experiments provide an opportunity for creating datasets that are
well-matched to 21cm observations from HERA and SKA. Unlike optical and near-IR
experiments which will have a relatively narrow field of view, LIM observations
promise widefield maps suitable for
cross-correlations. Additionally, LIM techniques incorporate spectroscopic
redshift information, which allows for meaningful comparison with 21 cm data
which has had foreground cleaning or avoidance techniques applied.
LIM-21 cm cross-correlations can help with constraining the properties of typical
high-redshift galaxies, due to their ability to be used in inferring the size distribution of ionized
regions during the EoR \citep{2009ApJ...690..252L,2011ApJ...741...70L}.\\

\noindent{\bf 21cm-CXB}\\
Once the fluctuations in the CXB are measured they
can be cross-correlated with the 21 cm measurements by experiments such as HERA
or the SKA. Properties of the first X-ray sources (including the bolometric
luminosity, spectral energy distribution, spatial distribution of these sources
and growth of the population in time)
strongly affect predictions for the
neutral hydrogen emission from redshifts $6 \lesssim z \lesssim 26$ with the exact range
depending on model parameters.
Cross-correlation between the CXB and the 21 cm signal is expected to be strong
over the period when the 21 cm signal is driven by X-ray heating (e.g., \citealt{2018MNRAS.480...26M}).
Moreover, the typical energy of produced radiation is imprinted in
the shape of the cross-correlation. Thus, properties of typical
high-redshift X-ray sources can be extracted from the
CXB and the 21 cm signals.

\subsection*{The Road Ahead}
\vspace{-7pt}

\begin{table}[t]
  \small
  \centering
  \begin{tabular}{|c|c|c|c|c|}
    \hline
    \textbf{Experiment} & \textbf{Tracer} & \textbf{Redshift} & \textbf{Angular Scales} & \textbf{Sky Coverage} \\ \hline
    HERA & 21 cm & $5 < z < 25$ & $11' < \theta < 120^\circ$ & 1440 deg$^2$ \\ \hline
    TIME & [CII] & $5 < z < 9$ & $30'' < \theta < 1^\circ$ & 390 arcmin$^2$ \\ \hline
    CONCERTO & [CII] & $4.25 < z < 8.5$ & $20'' < \theta < 1^\circ$ & 1.4 deg$^2$ \\ \hline
    COMAP & CO 2-1 & $5.8 < z < 7.8$ & $4' < \theta < 2^\circ$ & 5 deg$^2$ \\ \hline
    CCAT-p & [CII] & $3.25 < z < 9.25$ & $1' < \theta < 8^\circ$ & 64 deg$^2$ \\ \hline
    CDIM$^*$ & Ly$\alpha$ & $6 < z < 8$ & $1'' < \theta < 10^\circ$ & 100 deg$^2$ \\ \hline
    Simons Observatory & CMB & $z = 1100$ & $1' < \theta < 8^\circ$ & 16500 deg$^2$ \\ \hline
    Lynx$^*$ & X-ray & $z \lesssim 10$ & $ 0.3'' < \theta < 1^\circ$ & 1 deg$^2$ \\ \hline
    SKA-Low Phase 1 & 21 cm & $5 < z < 25$ & $5' < \theta < 20^\circ$ & 100 deg$^2$ \\ \hline
  \end{tabular}
  \caption{A summary of the observational capabilities of several
    upcoming experiments which will make maps relevant to reionization physics. CDIM and Lynx
    are proposed experiments, and so the final observational strategies are not yet fixed.}
  \label{table:experiments}
\end{table}

While the science opportunities summarized above are potentially transformative
significant challenges remain. Below we list our key recommendations for the coming decade. These recommendations focus on leveraging the full power of cross-correlations between maps generated by different experiments. 
\\

\noindent{\bf Recommendation 1: Promote Imaging Cross-correlation Studies}\\
As a first step toward cross-correlation measurements, communication between the
survey patterns and strategies for different experiments will be crucial for
providing as much suitable data as possible. Beyond this, sharing
data between experiments can be
facilitated by the development of robust community-driven software libraries
(e.g., \citealt{2018AJ....156..123A,Hazelton:2017:PPI}). Finally, efficient methods for
generating maps and computing the statistical estimators required to extract the signal from these
maps are an active and ongoing area of research. In particular, higher-point
estimators such as the bispectrum and trispectrum can extract the non-Gaussian
information present in the field. Other image-based methods can be used to
implicitly leverage all non-Gaussian information present in maps
\citep{2018arXiv181008211L}.\\

\noindent{\bf Recommendation 2: Develop Self-Consistent Large-Scale Simulations}\\
During cosmic dawn and the EoR, the
detailed astrophysics of individual stars is indirectly coupled to the large
scale observables in the
IGM. Realistically and self-consistently modeling these tremendously different
scales remains a theoretical challenge. Degeneracies in these physical models can be broken in principle by observations using different tracers, to get a more complete picture of the evolution of the Universe.
Figure~\ref{fig:kspace} and Table~\ref{table:experiments}
sketch the coverage of Fourier modes, angular scales, and redshift ranges accessed by various current and planned experiments
that will make maps of the EoR. These experiments will span many orders of
magnitude, from proto-galaxy cluster scales (100 kpc) up to truly cosmological
volumes (1 Gpc). Techniques for including predictions for the various physical tracers accurately for all scales of interest is highly non-trivial, as well as how to best incorporate observational artifacts of different experiments.


\newpage

\bibliographystyle{aasjournal}
\bibliography{references}


\end{document}